\documentclass[aps,secnumarabic,nobalancelastpage,amsmath,amssymb,
nofootinbib,preprint]{revtex4}
\usepackage{graphicx}
\usepackage{graphics}
\usepackage{amsmath}
\usepackage{dcolumn}
\usepackage{amssymb}
\usepackage{bm}
\usepackage{longtable}   
\usepackage{url} 
\usepackage{hyperref}
\begin{document}
\title{Dirac ocillator in the cosmic string spacetime in the context of gravity's rainbow}
\author{K. Bakke}
\email{kbakke@fisica.ufpb.br}
\affiliation{Departamento de F\'isica, Universidade Federal da Para\'iba, Caixa Postal 5008, 58051-900, Jo\~ao Pessoa, PB, Brazil.}

\author{H. Mota}
\email{hmota@fisica.ufpb.br}
\affiliation{Departamento de F\'isica, Universidade Federal da Para\'iba, Caixa Postal 5008, 58051-900, Jo\~ao Pessoa, PB, Brazil.}

\begin{abstract}
In this paper we consider the Dirac oscillator in the context of Doubly General Relativity or Gravity's Rainbow. In order to obtain the energy levels of the Dirac oscillator, we solve the Dirac equation in the cosmic string spacetime modified by gravity's rainbow scenarios described by two rainbow functions. We then obtain that, as a consequence of the modification of the cosmic string line element  by the two rainbow functions, the energy levels of the Dirac oscillator are appreciable altered. The results are plotted and compared with the standard case, without gravity's rainbow effects.


\end{abstract}
\maketitle

\section{Introduction}
Semi-classical approaches to investigate quantum gravity phenomena are enormously useful, and any achievement coming from those should be expected to be present in the not yet known fully-fledged theory of quantum gravity. In this sense, an interesting phenomenological framework to study quantum gravity effects is the one of Doubly General Relativity or, as it is mostly known, Gravity's Rainbow \cite{Magueijo:2002am,Magueijo:2002xx, AmelinoCamelia:2008qg}. In this framework, the definition of a nontrivial dual space, as a consequence of a nonlinear Lorentz transformation in momentum space, implies that the metric describing the spacetime should be energy-dependent, leading to a modification of the relativistic dispersion relation. This energy is due to the particle(s) which ultimately probes the spacetime since for each value of the particle frequency it feels a different geometry.

In the context of Gravity's Rainbow, there are in fact two quantities which are observer-independent (invariant): the speed of light and the Planck length (energy) \cite{AmelinoCamelia:2000mn,Magueijo:2001cr, Galan:2004st}.  Therefore, at high-energy scales (of the order of the Planck scale) the relativistic dispersion relation should acquire corrections \cite{Jacob:2010vr, AmelinoCamelia:1997gz}. This is in accordance, for instance, with ultra-high energy cosmic rays and TeV photons phenomena detected in experiments, suggesting the need of a modification of the relativistic dispersion relation \cite{AmelinoCamelia:1997gz,Magueijo:2002xx,AmelinoCamelia:2000mn, AmelinoCamelia:2008qg}.

The semi-classical approach of Gravity's Rainbow has been used, for instance, to investigate Cosmological and Astrophysical phenomena in a variety of contexts ranging from Friedmann-Robertson-Walker Universe \cite{Khodadi:2016bcx, Awad:2013nxa, Majumder:2013ypa, Hendi:2017vgo},  black hole thermodynamics \cite{Hendi:2016dmh, Hendi:2015cra, Hendi:2015bba, Hendi:2016vux, Hendi:2016njy}  to neutron stars' properties \cite{Hendi:2015vta}, massive scalar field in the Schwarzschild metric  \cite{Leiva:2008fd, Li:2008gs, Bezerra:2017hrb} and Casimir effect \cite{Bezerra:2017zqq}. These works show us a growing interest in the semi-classical approach of Rainbow's Gravity as to investigate quantum gravity effects through corrections to the dispersion relation of relativistic quantum fields. Hence, in this work, we raise a discussion within the context of gravity's rainbow by means of the effects that stem from high-order corrections to the relativistic dispersion relation of the Dirac oscillator in the cosmic string spacetime. The Dirac oscillator \cite{Moshinsky,Villalba:1993fq} is a relativistic model for the well-known harmonic oscillator \cite{Landau, Griffiths}, which is characterized by a coupling that keeps the Dirac equation being linear in both spatial coordinates and momenta. Besides, in the nonrelativistic limit of the Dirac equation, it recovers the spectrum of energy of the harmonic oscillator with a strong spin-orbit coupling. It has inspired a great deal of work \cite{Rozmej:1999jv, Boumali2013, Quesne:2004pp, 2008PhRvA..77c3832B, Karwowski2007, Mandal:2009we, PhysRevA.84.032109, Bakke:2013wla, Bakke:2012zz, Bakke:2013sla, Bakke:2011qr, Hassanabadi:2015hxa}. Therefore, our interest is to extend the discussion about the semi-classical approach of gravity's by analysing the influence of backgrounds determined by gravity's rainbow on the spectrum of energy of the Dirac oscillator.


The structure of this paper is: in Sec.2 we introduce the cosmic string spacetime and the essential aspects of gravity's rainbow framework that will modify the spacetime considered. Next, we solve the Dirac equation in the modified cosmic string spacetime, obtain the energy levels in two gravity's rainbow scenarios presented and discuss the results comparing it with the results for the energy levels of the Dirac oscillator without gravity's rainbow considered previously. Finally, in Sec.3 we present the conclusions. Throughout the paper we use natural units $G = \hbar = c = 1.$
%
\section{Dirac equation in the cosmic string gravity's rainbow}
In this section, we study the behaviour of the Dirac oscillator in two gravity's rainbow scenarios, which we shall specify latter. Firstly, we start by introducing the spacetime background that we wish to work with, namely, the cosmic string spacetime described by the following line element \cite{VS,hindmarsh}: 
\begin{eqnarray}
ds^{2}=-dt^{2}+dr^{2}+\eta^{2}r^{2}d\varphi^{2}+dz^{2},
\label{1.1}
\end{eqnarray}
where the parameter $\eta$ is related to the deficit angle which is defined as $\eta=1-4\varpi$, with $\varpi$ being the linear mass density of the cosmic string. In the cosmic string spacetime, we have that $\eta<1$ \cite{Katanaev:1992kh,Furtado:1994nq}. Furthermore, in the cylindrical symmetry we have that $0\,<\,r\,<\,\infty$, $0\leq\varphi\leq2\pi$ and $-\infty\,<\,z\,<\,\infty$. Cosmic string is a topological defect characterized by a spacetime with a conical topology and may have been produced by phase transitions in the early universe as it is predicted in extensions of the Standard Model of Particle Physics \cite{VS,hindmarsh}. Cosmic strings are also predicted in the framework of String Theory \cite{Hindmarsh:2011qj}. Once formed, cosmic strings can evolve in the Universe and contribute to a variety of astrophysical, cosmological and gravitational phenomena \cite{VS,hindmarsh,escidoc:153364, Mota:2014uka}, making the physics associated to them being of great relevance.

Let us now consider the framework of gravity's rainbow. In the latter, as it has been said previously, the high-energy scale regime dictates that the relativistic dispersion relation, associated with a given quantum field of mass $m$ and momentum $p$, has to be modified according to \cite{Magueijo:2002am,Magueijo:2002xx,Bezerra:2017zqq}
\begin{eqnarray}
E^{2}\,g_{0}^{2}\left(x\right)-p^{2}\,g_{1}^{2}\left(x\right)=m^{2}.
\label{1.2}
\end{eqnarray} 
The functions $g_{0}\left(x\right)$ and $g_{1}\left(x\right)$ are called rainbow functions and their argument, $x=E/E_{P}$, is the ratio of the energy of the probe particle to the Planck energy $E_{P}$. This ratio regulates the level of the mutual relation between the spacetime background and the probe particles. Therefore, at low-energy regimes we must have 
\begin{equation}
\lim_{x\rightarrow 0} g_{i}(x) = 1, \qquad \text{with} \qquad i=0,1.
\end{equation}
It should be mentioned that it is possible to get gravity's rainbow scenarios from theories with non-commutative geometries or other theories for quantum gravity such as loop quantum gravity \cite{Ashour:2016cay, 2016arXiv160806824F, Alsaleh:2017oae}. It means that the rainbow functions can also be derived from these theories.

In the context of gravity's rainbow, the line element of the cosmic string (\ref{1.1}) becomes
\begin{eqnarray}
ds^{2}=-\frac{1}{g_{0}^{2}\left(x\right)}\,dt^{2}+\frac{1}{g_{1}^{2}\left(x\right)}\left[dr^{2}+\eta^{2}r^{2}d\varphi^{2}+dz^{2}\right].
\label{1.a}
\end{eqnarray}
This modification of the cosmic string spacetime by gravity's rainbow has been analysed previously in Ref. \cite{Momeni:2017cvl}.

In the next (sub-)section we shall consider the Dirac oscillator in the background described by the line element \eqref{1.a} in order to see how the resulting energy levels associated with it get modified by the considered rainbow functions. They should be a generalization of the energy levels associated with the Dirac oscillator obtained in the background given by the line element \eqref{1.1}. As it is known \cite{Bakke:2013wla}, the energy levels of the Dirac oscillator obtained in the background of the cosmic string are given by the expression
%
%
%
\begin{figure}[!htb]
\begin{center}
\includegraphics[width=0.4\textwidth]{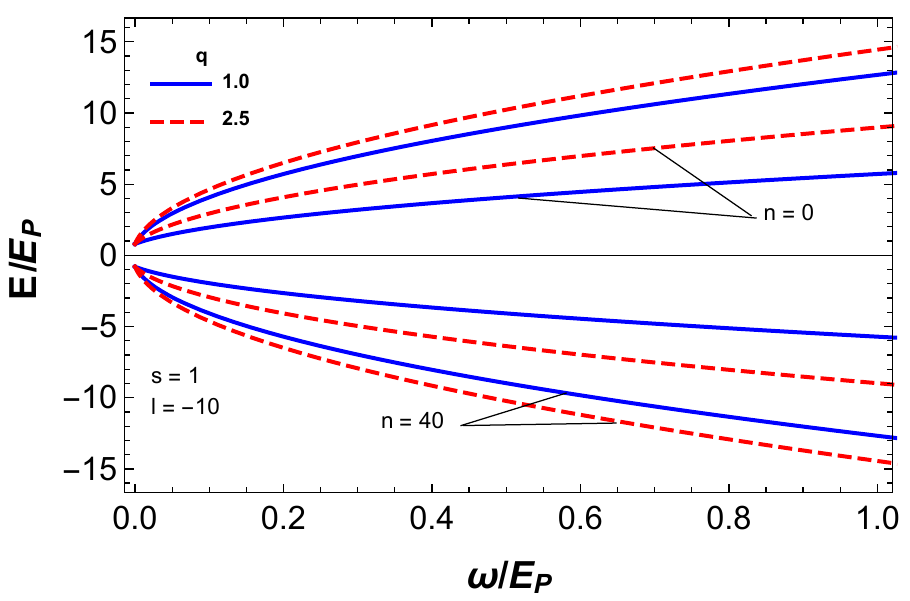}
%
\includegraphics[width=0.4\textwidth]{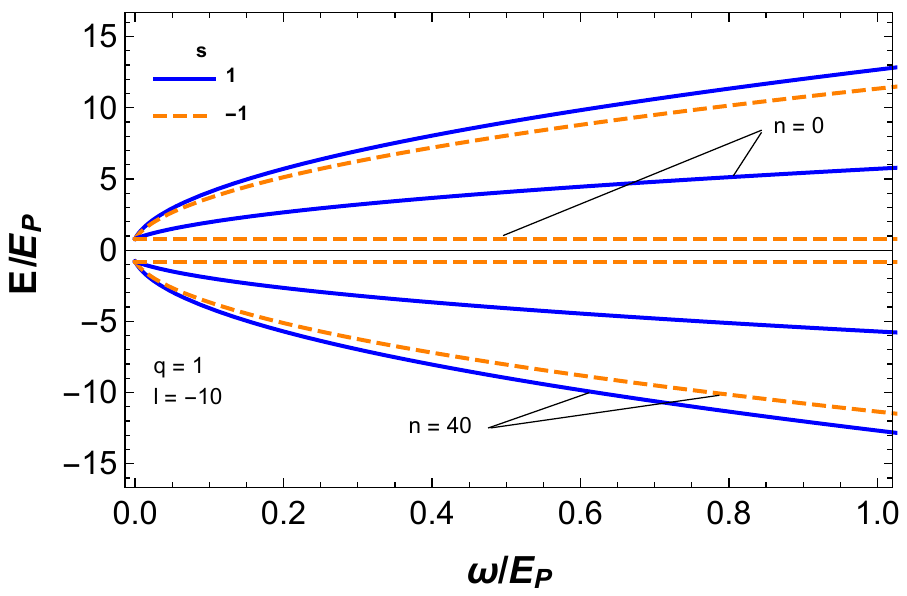}
\caption{\small{Plot of the energy levels \eqref{1.23} of the Dirac oscillator in the cosmic string spacetime, in units of the Planck energy, in terms of the ratio of the frequency $\omega$ to the Planck energy. For both plots we consider $\frac{m}{E_P}=0.8$ and $\eta = \frac{1}{q}$.}}
\label{f1}
\end{center}
\end{figure}
%
\begin{eqnarray}
E_{\sigma}=\pm\sqrt{m^{2}+4m\omega\left[n+\frac{\left|\nu\right|}{2\,\eta}-s\frac{\nu}{2\eta}\right]},
\label{1.23}
\end{eqnarray}
where $\omega$ is the frequency of the Dirac oscillator, $m$ is the mass and $\sigma = (n,l,s)$ is the set of quantum numbers which take the values: $n=0,1,2,...$, $l=0,\pm 1, \pm 2,...$ and $s=\pm 1$. The parameter $\nu$ is given by
\begin{equation}
\nu = l+\frac{1}{2}\left(1-s\right)+\frac{s}{2}\left(1-\eta\right).
\label{nu}
\end{equation}
Therefore, we can see that the energy levels above depends not only on the quantum numbers, mass, frequency but also on the cosmic string spacetime parameter $\eta$. In the limit $\eta\rightarrow 1$, one recovers the Dirac oscillator in Minkowski spacetime \cite{Bakke:2013wla}.

The expression in Eq. \eqref{1.23} is plotted in Fig.\ref{f1}, in units of the Planck energy, in terms of the ratio of the Dirac oscillator frequency to the Planck energy. For both plots we consider $\frac{m}{E_P} = 0.8$. The scale in which these energy levels are plotted is more appropriate for latter comparison with the ones from gravity's rainbow scenarios. On the left, the plot shows symmetric curves for values of the parameters indicated in the figure. The two internal curves (solid-blue and dashed-red) are for $n=0$ while the two external ones (solid-blue and dashed-red) are for $n=40$. It is clear that if we decrease the cosmic string parameter $\eta=\frac{1}{q}$, the positive and negative energies increase and decrease, respectively. On the other hand, the plot on the right shows symmetric curves for other set of parameter values as indicated in the figure. The dashed-orange and solid-blue curves are for different values of the spin, $s=\pm 1$. We can see that, for the same values of $n$ the energy gets bigger or smaller depending on whether we take $s=1$ or $s=-1$, respectively. One could take different values for the quantum numbers $n$ and $l$ but this would only increase or decrease the energy, keeping the same shape for the graph. Note that although the plots may lead the reader to believe that the energy levels go to zero when the frequency goes also to zero, that is not true. In fact, the energy gets a nonzero values when $\omega = 0$.
\subsection{First scenario of gravity's rainbow}
Let us now construct a scenario of the cosmic string gravity's rainbow by using the following rainbow functions \cite{Bezerra:2017zqq,Khodadi:2016bcx, Awad:2013nxa,Hendi:2017vgo}:
\begin{eqnarray}
g_{0}\left(x\right)=g_{1}\left(x\right)=\frac{1}{1-\epsilon x},
\label{1.3}
\end{eqnarray}
where we shall consider $\epsilon$ a first order parameter for our purposes. These rainbow functions fit the requirement of a constant velocity of light and solves the horizon problem puzzle \cite{Bezerra:2017zqq,Khodadi:2016bcx, Awad:2013nxa,Hendi:2017vgo}. In this first scenario of gravity's rainbow, described by the rainbow functions in Eq. \eqref{1.3}, the cosmic string line element \eqref{1.1} is modified according to \eqref{1.a} as 
\begin{eqnarray}
ds^{2}=-\left(1-\epsilon x\right)^{2}\,dt^{2}+\left(1-\epsilon x\right)^{2}\left[dr^{2}+\eta^{2}\,r^{2}\,d\varphi^{2}+dz^{2}\right].
\label{1.4}
\end{eqnarray}
Thus, we can note that by taking $\epsilon\rightarrow0$ we recover the cosmic string line element (\ref{1.1}), as it should be.

 We can go further by using the spinor theory in curved space \cite{birrell1984quantum}. In this way, spinors are defined locally by introducing a noncoordinate basis $\hat{\theta}^{a}=e^{a}_{\,\,\,\mu}\left(x\right)\,dx^{\mu}$, whose components $e^{a}_{\,\,\,\mu}\left(x\right)$ are called tetrads and give rise to the local reference frame of the observers. The tetrads satisfy the following relation:
\begin{eqnarray}
g_{\mu\nu}\left(x\right)=e^{a}_{\,\,\,\mu}\left(x\right)e^{b}_{\,\,\,\nu}\left(x\right)\,\eta_{ab},
\label{1.5}
\end{eqnarray}
where $\eta_{ab}=\mathrm{diag}\left(-\,+\,+\,+\right)$ is the Minkowski tensor. We also have the inverse of the tetrads, which is given by $dx^{\mu}=e^{\mu}_{\,\,\,a}\left(x\right)\,\hat{\theta}^{a}$, with $e^{a}_{\,\,\,\mu}\left(x\right)e^{\mu}_{\,\,\,b}\left(x\right)=\delta^{a}_{b}$. Thereby, from the modified line element of the cosmic string spacetime given in Eq. (\ref{1.4}), we can write the tetrads as
\begin{eqnarray}
\hat{\theta}^{0}=\left(1-\epsilon x\right)dt;\,\,\,\hat{\theta}^{1}=\left(1-\epsilon x\right)dr;\,\,\,\hat{\theta}^{2}=\eta\,r\left(1-\epsilon x\right)d\varphi;\,\,\,\hat{\theta}^{3}=\left(1-\epsilon x\right)dz.
\label{1.6}
\end{eqnarray}
Then, by solving the Maurer-Cartan structure equations in the absence of torsion \cite{nakahara2003geometry}, $d\hat{\theta}^{a}+\omega^{a}_{\,\,\,b}\wedge\hat{\theta}^{b}$ (where $\omega^{a}_{\,\,\,b}=\omega^{\,\,\,a}_{\mu\,\,\,b}\left(x\right)\,dx^{\mu}$), we obtain $\omega_{\varphi\,2\,1}\left(x\right)=-\omega_{\varphi\,1\,2}\left(x\right)=\eta$. Thus, the spinorial connection $\Gamma_{\mu}\left(x\right)=\frac{i}{4}\,\omega_{\mu ab}\left(x\right)\,\Sigma^{ab}$ ($\Sigma^{ab}=\frac{i}{2}\left[\gamma^{a},\gamma^{b}\right]$) has the following component
\begin{eqnarray}
\Gamma_{\varphi}\left(x\right)=-\frac{i}{2}\,\eta\,\Sigma^{3}.
\label{1.7}
\end{eqnarray}
Note that we have defined the $\gamma^{a}$ matrices in the local reference frame, where they correspond to the Dirac matrices in the Minkowski spacetime \cite{greiner1990relativistic,Bjorken}:
\begin{eqnarray}
\gamma^{0}=\left(
\begin{array}{cc}
1 & 0 \\
0 & -1 \\
\end{array}\right);\,\,\,\,\,\,
\gamma^{i}=\left(
\begin{array}{cc}
 0 & \sigma^{i} \\
-\sigma^{i} & 0 \\
\end{array}\right);\,\,\,\,\,\,\Sigma^{i}=\left(
\begin{array}{cc}
\sigma^{i} & 0 \\
0 & \sigma^{i} \\	
\end{array}\right),
\label{1.8}
\end{eqnarray}
with $\vec{\Sigma}$ as being the spin vector. The matrices $\sigma^{i}$ are the Pauli matrices and satisfy the relation $\frac{1}{2}\left(\sigma^{i}\,\sigma^{j}+\sigma^{j}\,\sigma^{i}\right)=\eta^{ij}$. The $\gamma^{\mu}$ matrices are related to the $\gamma^{a}$ matrices via $\gamma^{\mu}=e^{\mu}_{\,\,\,a}\left(x\right)\gamma^{a}$ \cite{birrell1984quantum}.

The Dirac oscillator \cite{Moshinsky,Villalba:1993fq} is a relativistic model for the harmonic oscillator, which is introduced into the Dirac equation through the coupling $\vec{p}\rightarrow\vec{p}-im\omega_{0}\,r\,\gamma^{0}\,\hat{r}$. With this coupling, the Dirac equation remains linear in both spatial coordinates and momenta. Therefore, the covariant form of the Dirac equation for this relativistic quantum oscillator is   
\begin{eqnarray}
i\gamma^{\mu}\partial_{\mu}\psi+i\gamma^{\mu}\Gamma_{\mu}\left(x\right)\psi+i\,m\,\omega\,\gamma^{\mu}\,X_{\mu}\,\gamma^{0}\psi=m\psi,
\label{1.9}
\end{eqnarray}
where $X_{\mu}=\left(0,r,0,0\right)$. In this way, by using (\ref{1.6}), (\ref{1.7}) and (\ref{1.8}), the Dirac equation for the Dirac oscillator becomes ($\hbar=c=1$)
\begin{eqnarray}
i\frac{\partial\psi}{\partial t}=m\,\left(1-\epsilon x\right)\,\gamma^{0}\psi-i\gamma^{0}\gamma^{1}\left(\frac{\partial}{\partial r}+\frac{1}{2r}+m\omega r\,\gamma^{0}\right)\psi-i\frac{\gamma^{0}\gamma^{2}}{\eta\,r}\,\frac{\partial\psi}{\partial\varphi}-i\gamma^{0}\gamma^{3}\,\frac{\partial\psi}{\partial z}.
\label{1.10}
\end{eqnarray}
The solution to the Dirac equation (\ref{1.10}) is given in the form: 
\begin{eqnarray}
\psi=e^{-i\,E\,t}\,\left(
\begin{array}{c}
\phi_{1}\\
\phi_{2}\\
\end{array}\right),
\label{1.11}
\end{eqnarray}
where $\phi_{1}$ and $\phi_{2}$ are spinors of two-components \cite{Bakke:2013wla}. Then, we obtain two coupled equations for $\phi_{1}$ and $\phi_{2}$, where the first one is
\begin{eqnarray}
\left[E-m\,\left(1-\epsilon x\right)\right]\phi_{1}=-i\sigma^{1}\left[\frac{\partial}{\partial r}+\frac{1}{2r}-m\omega r\right]\phi_{2}-i\frac{\sigma^{2}}{\eta\,r}\,\frac{\partial\phi_{2}}{\partial\varphi}-i\sigma^{3}\,\frac{\partial\phi_{2}}{\partial z},
\label{1.12}
\end{eqnarray}
while the second coupled equation is
\begin{eqnarray}
\left[E+m\,\left(1-\epsilon x\right)\right]\phi_{2}=-i\sigma^{1}\left[\frac{\partial}{\partial r}+\frac{1}{2r}+m\omega r\right]\phi_{1}-i\frac{\sigma^{2}}{\eta\,r}\,\frac{\partial\phi_{1}}{\partial\varphi}-i\sigma^{3}\,\frac{\partial\phi_{1}}{\partial z}.
\label{1.13}
\end{eqnarray}

By eliminating $\phi_{2}$ in Eqs. (\ref{1.12}) and (\ref{1.13}), therefore, we obtain the following equation for $\phi_{1}$:
\begin{eqnarray}
\left[E^{2}-m^{2}\left(1-\epsilon x\right)^{2}\right]\phi_{1}&=&-\frac{\partial^{2}\phi_{1}}{\partial r^{2}}-\frac{1}{r}\frac{\partial\phi_{1}}{\partial r}-\frac{1}{\eta^{2}r^{2}}\frac{\partial^{2}\phi_{1}}{\partial\varphi^{2}}-\frac{\partial^{2}\phi_{1}}{\partial z^{2}}+\frac{i\sigma^{3}}{\eta\,r^{2}}\frac{\partial\phi_{1}}{\partial\varphi}+\frac{1}{4\,r^{2}}\phi_{1}\nonumber\\
&+&m^{2}\omega^{2}\,r^{2}\,\phi_{1}-m\omega\,\phi_{1}+\frac{2m\omega}{\eta}\,i\sigma^{3}\,\frac{\partial\phi_{1}}{\partial\varphi}-2m\omega\,r\,i\sigma^{2}\,\frac{\partial\phi_{1}}{\partial z}.
\label{1.14}
\end{eqnarray}

Observe that $\sigma^{3}\phi_{1}=\pm\phi_{1}=s\phi_{1}$, where $s=\pm1$ and $\phi_{1}=\left(f_{+}\,\,\,f_{-}\right)^{T}$. Furthermore, due to the cylindrical symmetry of system, we can write $\phi_{1}=e^{i\left(l+1/2\right)\varphi}\,e^{i\,p_{z}\,z}\,\left(f_{+}\left(r\right)\,\,\,f_{-}\left(r\right)\right)^{T}$, where $l=0,\pm1,\pm2,\ldots$ and $-\infty\,<\,p_{z}\,<\,\infty$. From now on, let us take $p_{z}=0$, hence, we obtain the following second order differential equation for both $f_{+}\left(r\right)$ and $f_{-}\left(r\right)$:
\begin{eqnarray}
f_{s}''+\frac{1}{r}\,f_{s}'-\frac{\nu^{2}}{\eta^{2}r^{2}}\,f_{s}-m^{2}\omega^{2}r^{2}\,f_{s}+\beta\,f_{s}=0,
\label{1.15}
\end{eqnarray}  
where we have defined the parameterss
\begin{eqnarray}
\beta&=&E^{2}-m^{2}\left(1-\epsilon x\right)^{2}+2s\,m\,\omega\,\frac{\gamma_{s}}{\eta}+2m\omega;\nonumber\\
[-2mm]\label{1.16}\\[-2mm]
\nu&=&l+\frac{1}{2}\left(1-s\right)+\frac{s}{2}\left(1-\eta\right).\nonumber
\end{eqnarray}

By defining $\xi=m\omega\,r^{2}$, we obtain the following equation 
\begin{eqnarray}
\xi\,f_{s}''+f_{s}'-\frac{\nu^{2}}{4\eta^{2}\xi}\,f_{s}-\frac{\xi}{4}\,f_{s}+\frac{\beta}{4m\omega}\,f_{s}=0.
\label{1.17}
\end{eqnarray}
The solutions to this equation are given by
\begin{eqnarray}
f_{s}\left(\xi\right)=e^{-\frac{\xi}{2}}\,\,\,\xi^{\left|\nu\right|/2\eta}\,\,\,_{1}F_{1}\left(\frac{\left|\nu\right|}{2\eta}+\frac{1}{2}-\frac{\beta}{4m\omega},\,\,\frac{\left|\nu\right|}{\eta}+1;\,\,\xi\right),
\label{1.18}
\end{eqnarray}
where $\,_{1}F_{1}\left(\frac{\left|\nu\right|}{2\eta}+\frac{1}{2}-\frac{\beta}{4m\omega},\,\,\frac{\left|\nu\right|}{\eta}+1;\,\,\xi\right)$ is the confluent hypergeometric function \cite{Abramowitz,Arfken}. Observe that the asymptotic behaviour of the confluent hypergeometric function for large values of its argument is given by \cite{Abramowitz}
\begin{eqnarray}
\,_{1}F_{1}\left(a,\,b\,;x\right)\approx\frac{\Gamma\left(b\right)}{\Gamma\left(a\right)}\,e^{x}\,x^{a-b}\left[1+\mathcal{O}\left(\left|x\right|^{-1}\right)\right],
\label{1.19}
\end{eqnarray}
therefore, it diverges when $x\rightarrow\infty$. With the purpose of obtaining bound states solutions to the Dirac equation, we need to impose that $a=-n$ ($n=0,1,2,3,\ldots$), i.e., we need that $\frac{\left|\nu\right|}{2\eta}+\frac{1}{2}-\frac{\beta}{4m\omega}=-n$. With this condition, the confluent hypergeometric function becomes well-behaved when $x\rightarrow\infty$. Then, from the relation $\frac{\left|\nu\right|}{2\eta}+\frac{1}{2}-\frac{\beta}{4m\omega}=-n$, we obtain
\begin{eqnarray}
E^{2}+\frac{2\epsilon\,m^{2}}{E_{P}\left(1-\frac{\epsilon^{2}\,m^{2}}{E_{p}^{2}}\right)}\,E-\frac{\left(m^{2}+\lambda\right)}{\left(1-\frac{\epsilon^{2}\,m^{2}}{E_{p}^{2}}\right)}=0,
\label{1.20}
\end{eqnarray}
where $\lambda=4m\omega\left[n+\frac{\left|\nu\right|}{2\,\eta}-s\frac{\nu}{2\eta}\right]$. Observe that Eq. (\ref{1.20}) is a second degree algebraic equation for $E$. Therefore, the allowed energies of the system are
\begin{eqnarray}
\label{el1}
E_{\sigma}&=&-\frac{\epsilon\,m^{2}}{E_{p}\left(1-\frac{\epsilon^{2}\,m^{2}}{E_{p}^{2}}\right)}\nonumber\\
[-2mm]\label{1.21}\\[-2mm]
&\pm&\frac{1}{\left(1-\frac{\epsilon^{2}\,m^{2}}{E_{p}^{2}}\right)}\sqrt{\left(m^{2}+4m\omega\left[n+\frac{\left|\nu\right|}{2\,\eta}-s\frac{\nu}{2\eta}\right]\right)\cdot\left(1-\frac{\epsilon^{2}\,m^{2}}{E_{p}^{2}}\right)+\frac{\epsilon^{2}m^{4}}{E_{p}^{2}}}.\nonumber
\end{eqnarray}
%
%
%
\begin{figure}[!htb]
\begin{center}
\includegraphics[width=0.45\textwidth]{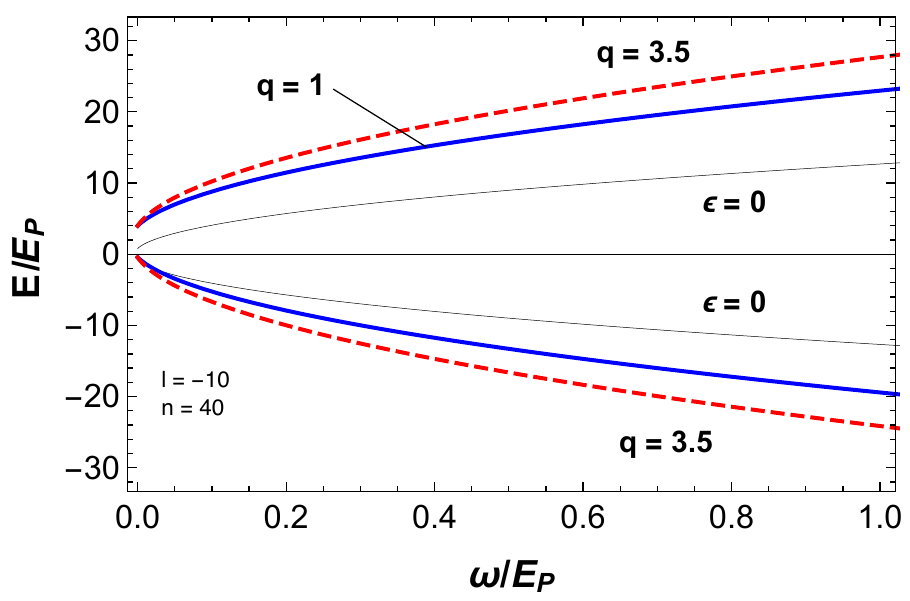}
\caption{\small{Plot of the energy levels \eqref{el1} of the Dirac oscillator in the cosmic string gravity's rainbow, in units of the Planck energy, in terms of the ratio of the frequency $\omega$ to the Planck energy. For this plot we consider $\frac{m}{E_P}=0.8$ and $\eta = \frac{1}{q}$. Note that the black thin curves (for $q=1$), describe the energy levels in the absence of gravity's rainbow while the color curves describe the energy leves with the gravity's rainbow modifications.}}
\label{f2}
\end{center}
\end{figure}
%
%

By comparing with the spectrum of energy \eqref{1.23}, obtained in Ref. \cite{Bakke:2013wla}, we can see that the modified background of the cosmic string spacetime changes the spectrum of energy of the Dirac oscillator by yielding the allowed energies (\ref{1.21}). The effects that stem from the topology of the cosmic string are given by the presence of the parameter $\eta$ in the allowed energies (\ref{1.21}). By taking the limit $\eta\rightarrow1$ in Eq. (\ref{1.21}), we have 
\begin{eqnarray}
\label{el2}
E_{\sigma}&=&-\frac{\epsilon\,m^{2}}{E_{p}\left(1-\frac{\epsilon^{2}\,m^{2}}{E_{p}^{2}}\right)}\nonumber\\
[-2mm]\label{1.22}\\[-2mm]
&\pm&\frac{1}{\left(1-\frac{\epsilon^{2}\,m^{2}}{E_{p}^{2}}\right)}\sqrt{\left(m^{2}+4m\omega\left[n+\frac{\left|\bar{\gamma}\right|}{2}-s\frac{\bar{\gamma}}{2}\right]\right)\cdot\left(1-\frac{\epsilon^{2}\,m^{2}}{E_{p}^{2}}\right)+\frac{\epsilon^{2}m^{4}}{E_{p}^{2}}}.\nonumber
\end{eqnarray}
where $\bar{\gamma}=l+\frac{1}{2}\left(1-s\right)$. In this case, we have a modified background of the Minkowski spacetime. Therefore, the effects of gravity's rainbow modifies the energy levels of the Dirac oscillator in contrast to that obtained in Refs. \cite{Moshinsky, Villalba:1993fq} in the Minkowski spacetime. 

In Fig.\ref{f2} we have plotted the energy levels \eqref{el1} of the Dirac oscillator in the cosmic string gravity's rainbow, in units of the Planck energy, in terms of the ratio of the frequency $\omega$ to the Planck energy. The values of the parameters considered are indicated on the figure. Also, for this plot we consider $\frac{m}{E_P}=0.8$ and $\eta = \frac{1}{q}$. Note that the black thin curves describe the energy levels, taking $q=1$, in the absence of gravity's rainbow while the color curves describe the energy leves with gravity's rainbow modifications. It is clear that, compared with the black thin curves (see also Fig.\ref{f1}), the energy leves described by the solid-blue curves (which is also for $q=1$) increases (decreases) in the cosmic string gravity's rainbow context. The same is true for the dashed-red line, which is for $q=2.5$. It is interesting to note that, the curves for the positive energy levels are shifted in the vertical direction when we considered gravity's rainbow scenario given by the rainbow functions \eqref{1.3}. Thereby, the symmetry shown in Fig.\ref{f1} between the curves for the positive and negative energy levels is lost. This happens because of the first term in Eqs. \eqref{el1} and \eqref{el2}. Note again that, for $\omega = 0$, the curves do not go to zero as we can verify from Eqs. \eqref{el1} and \eqref{el2}.
%
\subsection{Second scenario of gravity's rainbow}

In this section, we analyse the behaviour of the Dirac oscillator in another scenario of gravity's rainbow. Thus, the second scenario of gravity's rainbow we want to consider is the one defined by the rainbow functions \cite{Bezerra:2017zqq,Khodadi:2016bcx, Awad:2013nxa,Hendi:2017vgo}:
\begin{eqnarray}
g_{0}\left(x\right)=1;\,\,\,\,g_{1}\left(x\right)=\sqrt{1-\epsilon\,x^{2}}.
\label{2.1}
\end{eqnarray}
Gravity's rainbow scenario constructed with the rainbow functions \eqref{2.1} can be achieved as a limiting case from theories based on non-commutative geometries and loop quantum gravity \cite{Bezerra:2017zqq,Khodadi:2016bcx, Awad:2013nxa,Hendi:2017vgo}. These rainbow functions have also been considered in many different contexts to study the effects of gravity's rainbow on the Friedmann-Robertson-Walker Universes \cite{Khodadi:2016bcx, Awad:2013nxa, Majumder:2013ypa, Hendi:2017vgo}.

By considering the rainbow functions \eqref{2.1}, the line element of the cosmic string spacetime in the context of gravity's rainbow (\ref{1.a}) is given by
\begin{eqnarray}
ds^{2}=-\,dt^{2}+\frac{1}{\left(1-\epsilon x^{2}\right)}\left[dr^{2}+\eta^{2}\,r^{2}\,d\varphi^{2}+dz^{2}\right].
\label{2.2}
\end{eqnarray}
With the line element (\ref{2.2}), let us write the tetrads as
\begin{eqnarray}
\hat{\theta}^{0}=dt;\,\,\,\hat{\theta}^{1}=\frac{1}{\left(1-\epsilon x^{2}\right)^{1/2}}\,dr;\,\,\,\hat{\theta}^{2}=\frac{\eta\,r}{\left(1-\epsilon x^{2}\right)^{1/2}}\,\,d\varphi;\,\,\,\hat{\theta}^{3}=\frac{1}{\left(1-\epsilon x^{2}\right)^{1/2}}\,dz.
\label{2.3}
\end{eqnarray}
Then, by solving again the Maurer-Cartan structure equations in the absence of torsion \cite{nakahara2003geometry}, we also obtain $\omega_{\varphi\,2\,1}\left(x\right)=-\omega_{\varphi\,1\,2}\left(x\right)=\eta$ and the spinorial connection (\ref{1.7}). Hence, the Dirac equation becomes
\begin{eqnarray}
i\frac{\partial\psi}{\partial t}&=&m\,\gamma^{0}\psi-i\left(1-\epsilon x^{2}\right)^{1/2}\,\gamma^{0}\gamma^{1}\left(\frac{\partial}{\partial r}+\frac{1}{2r}+m\omega r\,\gamma^{0}\right)\psi-i\left(1-\epsilon x^{2}\right)^{1/2}\,\frac{\gamma^{0}\gamma^{2}}{\eta\,r}\,\frac{\partial\psi}{\partial\varphi}\nonumber\\
[-2mm]\label{2.4}\\[-2mm]
&-&i\left(1-\epsilon x^{2}\right)^{1/2}\gamma^{0}\gamma^{3}\,\frac{\partial\psi}{\partial z}.\nonumber
\end{eqnarray}

Next, by taking the solution to the Dirac equation (\ref{1.11}), we obtain the following coupled equations for $\phi_{1}$ and $\phi_{2}$
\begin{eqnarray}
\left[\frac{E-m}{\sqrt{1-\epsilon x^{2}}}\right]\phi_{1}=-i\sigma^{1}\left[\frac{\partial}{\partial r}+\frac{1}{2r}-m\omega r\right]\phi_{2}-i\frac{\sigma^{2}}{\eta\,r}\,\frac{\partial\phi_{2}}{\partial\varphi}-i\sigma^{3}\,\frac{\partial\phi_{2}}{\partial z},
\label{2.5}
\end{eqnarray}
while the second coupled equation is
\begin{eqnarray}
\left[\frac{E+m}{\sqrt{1-\epsilon x^{2}}}\right]\phi_{2}=-i\sigma^{1}\left[\frac{\partial}{\partial r}+\frac{1}{2r}+m\omega r\right]\phi_{1}-i\frac{\sigma^{2}}{\eta\,r}\,\frac{\partial\phi_{1}}{\partial\varphi}-i\sigma^{3}\,\frac{\partial\phi_{1}}{\partial z}.
\label{2.6}
\end{eqnarray}
By eliminating $\phi_{2}$ in Eqs. (\ref{2.5}) and (\ref{2.6}), therefore, we obtain the following equation for $\phi_{1}$:
\begin{eqnarray}
\left[\frac{E^{2}-m^{2}}{\left(1-\epsilon\,x^{2}\right)}\right]\phi_{1}&=&-\frac{\partial^{2}\phi_{1}}{\partial r^{2}}-\frac{1}{r}\frac{\partial\phi_{1}}{\partial r}-\frac{1}{\eta^{2}r^{2}}\frac{\partial^{2}\phi_{1}}{\partial\varphi^{2}}-\frac{\partial^{2}\phi_{1}}{\partial z^{2}}+\frac{i\sigma^{3}}{\eta\,r^{2}}\frac{\partial\phi_{1}}{\partial\varphi}+\frac{1}{4\,r^{2}}\,\phi_{1}\nonumber\\
&+&m^{2}\omega^{2}\,r^{2}\,\phi_{1}-m\omega\,\phi_{1}+\frac{2m\omega}{\eta}\,i\sigma^{3}\,\frac{\partial\phi_{1}}{\partial\varphi}-2m\omega\,r\,i\sigma^{2}\,\frac{\partial\phi_{1}}{\partial z}
\label{2.7}
\end{eqnarray}

Hence, by following the steps from Eq. (\ref{1.14}) to Eq. (\ref{1.20}), we obtain 
\begin{eqnarray}
E_{\sigma}=\pm\sqrt{\frac{m^{2}+4m\omega\left[n+\frac{\left|\nu\right|}{2\,\eta}-s\frac{\nu}{2\eta}\right]}{\left(1+\frac{4m\omega\epsilon}{E_{P}^{2}}\,\left[n+\frac{\left|\nu\right|}{2\,\eta}-s\frac{\nu}{2\eta}\right]\right)}},
\label{2.8}
\end{eqnarray}
which is the spectrum of energy of the Dirac oscillator in the modified cosmic string spacetime (\ref{2.2}). In contrast to the energy levels of the Dirac oscillator in the cosmic string spacetime obtained in Ref. \cite{Bakke:2013wla}, we have a different spectrum of energy of the Dirac oscillator yielded by the effects of gravity's rainbow scenario determined by the rainbow functions (\ref{2.1}). We can also see that this spectrum of energy differs from that obtained in Eq. (\ref{1.21}). Hence, distinct scenarios of the rainbow gravity yield different spectra of energy of the Dirac oscillator. Observe in Eq. (\ref{2.8}) that the effects of the topology of the cosmic string are also given by the presence of the parameter $\eta$. By taking the limit $\eta\rightarrow1$ in Eq. (\ref{2.8}), we have the modified background of the Minkowski spacetime, and thus we obtain  
\begin{eqnarray}
E_{\sigma}=\pm\sqrt{\frac{m^{2}+4m\omega\left[n+\frac{\left|\bar{\gamma}\right|}{2}-s\frac{\bar{\gamma}}{2}\right]}{\left(1+\frac{4m\omega\epsilon}{E_{P}^{2}}\,\left[n+\frac{\left|\bar{\gamma}\right|}{2}-s\frac{\bar{\gamma}}{2}\right]\right)}},
\label{2.9}
\end{eqnarray}
where $\bar{\gamma}=l+\frac{1}{2}\left(1-s\right)$. Therefore, we can also observe the effects of gravity's rainbow in the energy levels \eqref{2.9} of the Dirac oscillator in contrast to that obtained in Refs. \cite{Moshinsky,Villalba:1993fq} in the Minkowski spacetime. 
%
%
\begin{figure}[!htb]
\begin{center}
\includegraphics[width=0.45\textwidth]{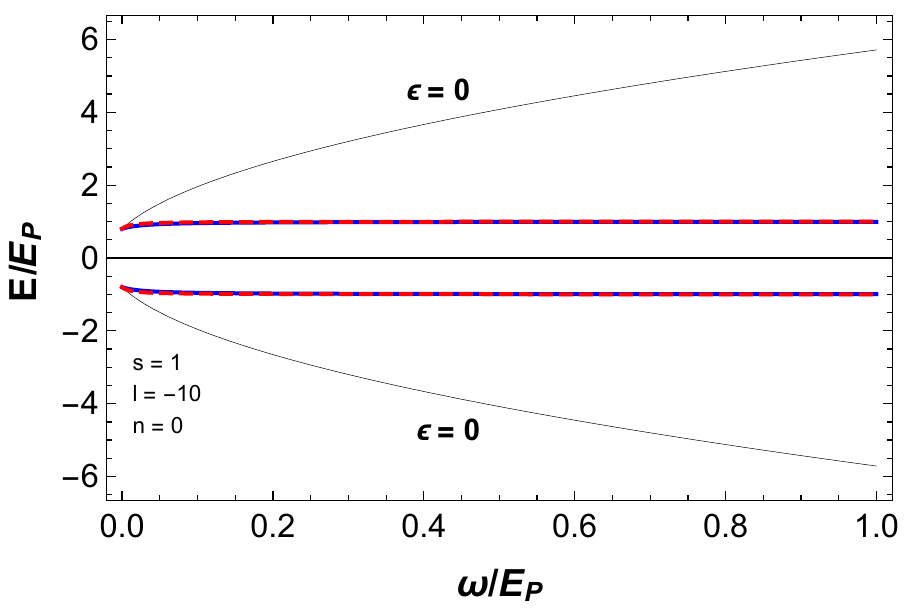}
\caption{\small{Plot of the energy levels \eqref{2.8} of the Dirac oscillator in the cosmic string gravity's rainbow, in units of the Planck energy, in terms of the ratio of the frequency $\omega$ to the Planck energy. For this plot we consider $\frac{m}{E_P}=0.8$ and and $\eta = \frac{1}{q}$. Note that the black thin curves (for $q=1$), describe the energy levels in the absence of gravity's rainbow while the colour curves (dashed-red for $q=2.5$ and solid-blue for $q=1$) describe the energy levels with gravity's rainbow modifications. }}
\label{f3}
\end{center}
\end{figure}
%
%
Finally, by taking $\epsilon\rightarrow0$ in Eq. (\ref{2.8}), we recover the energy levels of the Dirac oscillator in the cosmic string spacetime \cite{Bakke:2013wla}, i.e., we recover Eq. (\ref{1.23}).

In Fig.\ref{f3} we have plotted the energy levels \eqref{2.8} of the Dirac oscillator in the cosmic string gravity rainbow, in units of the Planck energy, in terms of the ratio of the frequency $\omega$ to the Planck energy. The values of the parameters considered are indicated on the figure. Also, for this plot we consider $\frac{m}{E_P}=0.8$ and and $\eta = \frac{1}{q}$. The conclusion here is similar to the one in Fig.\ref{f2}, that is, the black thin curves describe the energy levels, taking $q=1$, in the absence of gravity's rainbow while the colour curves describe the energy levels with gravity's rainbow modifications. Compared with the black thin curves (see also Fig.\ref{f1}), the energy levels described by the solid-blue curves (which is also for $q=1$) increase (decrease) in the cosmic string gravity rainbow context. The same is true for the dashed-red line, which is for $q=2.5$. Note that, increasing the cosmic string parameter $q$, the energy does not seem to change. Note also that, in this case, we do not have shifted curves for the positive energy levels, like in Fig.\ref{f2}. The reason for that is that we do not have a term like the first one in Eq. \eqref{el1}. Therefore, the curves shown in Fig.\ref{f3} preserve the symmetry between the positive and negative energy levels. Here is more evident that for $\omega = 0$, the curves do not go to zero.
\section{conclusions}
 We have analysed the Dirac oscillator in the cosmic string spacetime modified by two scenarios of gravity's rainbow. The first scenario is characterized by the rainbow function \eqref{1.3} while the second scenario is characterized by the rainbow function \eqref{2.1}, both of them considered in many different contexts. The line elements, \eqref{1.4} and \eqref{2.2}, describing the cosmic string gravity's rainbow, were then used to solve the Dirac equation, which made possible to obtain the modified energy levels \eqref{el1}-\eqref{el2} in the first scenario and \eqref{2.8}-\eqref{2.9} in the second scenario. We also pointed out that in the limit where gravity's rainbow parameter $\epsilon$ is taken to be zero, we recover the energy levels \eqref{1.23} that has been considered previously by the authors in \cite{Bakke:2013wla}.
 
 We have then plotted the energy levels \eqref{el1}-\eqref{el2} which are shown in Fig.\ref{f2} and compared it with the curves obtained from the energy levels \eqref{1.23} that is shown in Fig.\ref{f1}. The same is done with the plot of the energy levels \eqref{2.8}-\eqref{2.9} in Fig.\ref{f3}. In both cases we clearly saw that the curves for the energies are increased (decreased) for given values of the parameters involved in the problem. We emphasize that in the first case, shown in Fig.\ref{f2}, the curves for the positive part of the energy levels are shifted when compared with the standard case, without gravity's rainbow. Thus, the symmetry between the curves described by the negative and positive energy levels, as seen in Fig.\ref{f1}, is lost as indicated in Fig.\ref{f2}. 
\section*{Acknowledgments}
The authors would like to thank Prof. H. Hassanabadi for interesting discussions, and the Brazilian agency CNPq (Conselho Nacional de Desenvolvimento Cient\'{i}fico e Tecnol\'{o}gico - Brazil) for financial support.


\begin{thebibliography}{10}

\bibitem{Magueijo:2002am}
J.~Magueijo and L.~Smolin, {\it {Generalized Lorentz invariance with an
  invariant energy scale}},  {\em Phys. Rev.} {\bf D67} (2003) 044017,
  [\href{http://xxx.lanl.gov/abs/gr-qc/0207085}{{\tt gr-qc/0207085}}].

\bibitem{Magueijo:2002xx}
J.~Magueijo and L.~Smolin, {\it {Gravity's rainbow}},  {\em Class. Quant.
  Grav.} {\bf 21} (2004) 1725--1736,
  [\href{http://xxx.lanl.gov/abs/gr-qc/0305055}{{\tt gr-qc/0305055}}].

\bibitem{AmelinoCamelia:2008qg}
G.~Amelino-Camelia, {\it {Quantum-Spacetime Phenomenology}},  {\em Living Rev.
  Rel.} {\bf 16} (2013) 5, [\href{http://xxx.lanl.gov/abs/0806.0339}{{\tt
  arXiv:0806.0339}}].

\bibitem{AmelinoCamelia:2000mn}
G.~Amelino-Camelia, {\it {Relativity in space-times with short distance
  structure governed by an observer independent (Planckian) length scale}},
  {\em Int. J. Mod. Phys.} {\bf D11} (2002) 35--60,
  [\href{http://xxx.lanl.gov/abs/gr-qc/0012051}{{\tt gr-qc/0012051}}].

\bibitem{Magueijo:2001cr}
J.~Magueijo and L.~Smolin, {\it {Lorentz invariance with an invariant energy
  scale}},  {\em Phys. Rev. Lett.} {\bf 88} (2002) 190403,
  [\href{http://xxx.lanl.gov/abs/hep-th/0112090}{{\tt hep-th/0112090}}].

\bibitem{Galan:2004st}
P.~Galan and G.~A. Mena~Marugan, {\it {Quantum time uncertainty in a gravity's
  rainbow formalism}},  {\em Phys. Rev.} {\bf D70} (2004) 124003,
  [\href{http://xxx.lanl.gov/abs/gr-qc/0411089}{{\tt gr-qc/0411089}}].

\bibitem{Jacob:2010vr}
U.~Jacob, F.~Mercati, G.~Amelino-Camelia, and T.~Piran, {\it {Modifications to
  Lorentz invariant dispersion in relatively boosted frames}},  {\em Phys.
  Rev.} {\bf D82} (2010) 084021, [\href{http://xxx.lanl.gov/abs/1004.0575}{{\tt
  arXiv:1004.0575}}].

\bibitem{AmelinoCamelia:1997gz}
G.~Amelino-Camelia, J.~R. Ellis, N.~E. Mavromatos, D.~V. Nanopoulos, and
  S.~Sarkar, {\it {Tests of quantum gravity from observations of gamma-ray
  bursts}},  {\em Nature} {\bf 393} (1998) 763--765,
  [\href{http://xxx.lanl.gov/abs/astro-ph/9712103}{{\tt astro-ph/9712103}}].

\bibitem{Khodadi:2016bcx}
M.~Khodadi, K.~Nozari, and H.~R. Sepangi, {\it {More on the initial singularity
  problem in gravity?s rainbow cosmology}},  {\em Gen. Rel. Grav.} {\bf 48}
  (2016), no.~12 166, [\href{http://xxx.lanl.gov/abs/1602.0292}{{\tt
  arXiv:1602.0292}}].

\bibitem{Awad:2013nxa}
A.~Awad, A.~F. Ali, and B.~Majumder, {\it {Nonsingular Rainbow Universes}},
  {\em JCAP} {\bf 1310} (2013) 052,
  [\href{http://xxx.lanl.gov/abs/1308.4343}{{\tt arXiv:1308.4343}}].

\bibitem{Majumder:2013ypa}
B.~Majumder, {\it {Quantum Rainbow Cosmological Model With Perfect Fluid}},
  {\em Int. J. Mod. Phys.} {\bf D22} (2013), no.~13 1350079,
  [\href{http://xxx.lanl.gov/abs/1307.5273}{{\tt arXiv:1307.5273}}].

\bibitem{Hendi:2017vgo}
S.~H. Hendi, M.~Momennia, B.~Eslam~Panah, and S.~Panahiyan, {\it {Nonsingular
  Universe in Massive Gravity's Rainbow}},  {\em Universe} {\bf 16} (2017) 26,
  [\href{http://xxx.lanl.gov/abs/1705.0109}{{\tt arXiv:1705.0109}}].

\bibitem{Hendi:2016dmh}
S.~H. Hendi, S.~Panahiyan, B.~E. Panah, and M.~Momennia, {\it {Thermodynamic
  instability of nonlinearly charged black holes in gravity?s rainbow}},
  {\em Eur. Phys. J.} {\bf C76} (2016), no.~3 150,
  [\href{http://xxx.lanl.gov/abs/1512.0519}{{\tt arXiv:1512.0519}}].

\bibitem{Hendi:2015cra}
S.~H. Hendi, M.~Faizal, B.~E. Panah, and S.~Panahiyan, {\it {Charged dilatonic
  black holes in gravity?s rainbow}},  {\em Eur. Phys. J.} {\bf C76} (2016),
  no.~5 296, [\href{http://xxx.lanl.gov/abs/1508.0023}{{\tt arXiv:1508.0023}}].

\bibitem{Hendi:2015bba}
S.~H. Hendi, {\it {Asymptotically charged BTZ black holes in gravity?s
  rainbow}},  {\em Gen. Rel. Grav.} {\bf 48} (2016), no.~4 50,
  [\href{http://xxx.lanl.gov/abs/1507.0473}{{\tt arXiv:1507.0473}}].

\bibitem{Hendi:2016vux}
S.~H. Hendi, B.~Eslam~Panah, and S.~Panahiyan, {\it {Topological charged black
  holes in massive gravity's rainbow and their thermodynamical analysis through
  various approaches}},  {\em Phys. Lett.} {\bf B769} (2017) 191--201,
  [\href{http://xxx.lanl.gov/abs/1602.0183}{{\tt arXiv:1602.0183}}].

\bibitem{Hendi:2016njy}
S.~H. Hendi, S.~Panahiyan, B.~Eslam~Panah, M.~Faizal, and M.~Momennia, {\it
  {Critical behavior of charged black holes in Gauss-Bonnet gravity?s
  rainbow}},  {\em Phys. Rev.} {\bf D94} (2016), no.~2 024028,
  [\href{http://xxx.lanl.gov/abs/1607.0666}{{\tt arXiv:1607.0666}}].

\bibitem{Hendi:2015vta}
S.~H. Hendi, G.~H. Bordbar, B.~E. Panah, and S.~Panahiyan, {\it {Modified TOV
  in gravity's rainbow: properties of neutron stars and dynamical stability
  conditions}},  {\em JCAP} {\bf 1609} (2016), no.~09 013,
  [\href{http://xxx.lanl.gov/abs/1509.0514}{{\tt arXiv:1509.0514}}].

\bibitem{Leiva:2008fd}
C.~Leiva, J.~Saavedra, and J.~Villanueva, {\it {The Geodesic Structure of the
  Schwarzschild Black Holes in Gravity's Rainbow}},  {\em Mod. Phys. Lett.}
  {\bf A24} (2009) 1443--1451, [\href{http://xxx.lanl.gov/abs/0808.2601}{{\tt
  arXiv:0808.2601}}].

\bibitem{Li:2008gs}
H.~Li, Y.~Ling, and X.~Han, {\it {Modified (A)dS Schwarzschild black holes in
  Rainbow spacetime}},  {\em Class. Quant. Grav.} {\bf 26} (2009) 065004,
  [\href{http://xxx.lanl.gov/abs/0809.4819}{{\tt arXiv:0809.4819}}].

\bibitem{Bezerra:2017hrb}
V.~B. Bezerra, H.~R. Christiansen, M.~S. Cunha, and C.~R. Muniz, {\it {Exact
  solutions and phenomenological constraints from massive scalars in a
  gravity?s rainbow spacetime}},  {\em Phys. Rev.} {\bf D96} (2017), no.~2
  024018, [\href{http://xxx.lanl.gov/abs/1704.0121}{{\tt arXiv:1704.0121}}].

\bibitem{Bezerra:2017zqq}
V.~B. Bezerra, H.~F. Mota, and C.~R. Muniz, {\it {Casimir Effect in the Rainbow
  Einstein's Universe}},  {\em EPL} {\bf 120} (2017), no.~1 10005,
  [\href{http://xxx.lanl.gov/abs/1708.0262}{{\tt arXiv:1708.0262}}].

\bibitem{Moshinsky}
M.~Moshinsky and A.~Szczepaniak, {\it The dirac oscillator},  {\em Journal of
  Physics A: Mathematical and General} {\bf 22} (1989), no.~17 L817.

\bibitem{Villalba:1993fq}
V.~M. Villalba, {\it {Exact solution of the two-dimensional Dirac oscillator}},
   {\em Phys. Rev.} {\bf A49} (1994) 586,
  [\href{http://xxx.lanl.gov/abs/hep-th/9310010}{{\tt hep-th/9310010}}].

\bibitem{Landau}
L.~D. Landau and E.~M. Lifshitz, {\em Quantum Mechanics, the nonrelativistic
  theory, 3rd Ed.}
\newblock Pergamon, Oxford, 1977.

\bibitem{Griffiths}
L.~D. Landau and E.~M. Lifshitz, {\em D. J. Griffiths, Introduction to quantum
  mechanics, Second Edition}.
\newblock Prentice Hall, 2004.

\bibitem{Rozmej:1999jv}
P.~Rozmej and R.~Arvieu, {\it {The dirac oscillator. a relativistic version of
  the jaynes-cummings model}},  {\em J. Phys.} {\bf A32} (1999) 5367--5382,
  [\href{http://xxx.lanl.gov/abs/quant-ph/9903073}{{\tt quant-ph/9903073}}].

\bibitem{Boumali2013}
A.~Boumali and H.~Hassanabadi, {\it The thermal properties of a two-dimensional
  dirac oscillator under an external magnetic field},  {\em The European
  Physical Journal Plus} {\bf 128} (Oct, 2013) 124.

\bibitem{Quesne:2004pp}
C.~Quesne and V.~M. Tkachuk, {\it {Dirac oscillator with nonzero minimal
  uncertainty in position}},  {\em J. Phys.} {\bf A38} (2005) 1747--1766,
  [\href{http://xxx.lanl.gov/abs/math-ph/0412052}{{\tt math-ph/0412052}}].

\bibitem{2008PhRvA..77c3832B}
A.~{Bermudez}, M.~A. {Martin-Delgado}, and A.~{Luis}, {\it {Nonrelativistic
  limit in the 2+1 Dirac oscillator: A Ramsey-interferometry effect}},  {\em
  \pra} {\bf 77} (Mar., 2008) 033832,
  [\href{http://xxx.lanl.gov/abs/0709.2557}{{\tt arXiv:0709.2557}}].

\bibitem{Karwowski2007}
J.~Karwowski and G.~Pestka, {\it Harmonic oscillators in relativistic quantum
  mechanics},  {\em Theoretical Chemistry Accounts} {\bf 118} (Sep, 2007)
  519--525.

\bibitem{Mandal:2009we}
B.~P. Mandal and S.~Verma, {\it {Dirac oscillators in presence of external
  magnetic field}},  {\em Phys. Lett.} {\bf A374} (2010) 1021--1023,
  [\href{http://xxx.lanl.gov/abs/0907.4544}{{\tt arXiv:0907.4544}}].

\bibitem{PhysRevA.84.032109}
J.~Carvalho, C.~Furtado, and F.~Moraes, {\it Dirac oscillator interacting with
  a topological defect},  {\em Phys. Rev. A} {\bf 84} (Sep, 2011) 032109.

\bibitem{Bakke:2013wla}
K.~Bakke and C.~Furtado, {\it {On the interaction of the Dirac oscillator with
  the Aharonov-Casher system in topological defect backgrounds}},  {\em Annals
  Phys.} {\bf 336} (2013) 489--504,
  [\href{http://xxx.lanl.gov/abs/1307.2888}{{\tt arXiv:1307.2888}}].

\bibitem{Bakke:2012zz}
K.~Bakke, {\it {Noninertial effects on the Dirac oscillator in a topological
  defect spacetime}},  {\em Eur. Phys. J. Plus} {\bf 127} (2012) 82,
  [\href{http://xxx.lanl.gov/abs/1209.0369}{{\tt arXiv:1209.0369}}].

\bibitem{Bakke:2013sla}
K.~Bakke, {\it {Rotating effects on the Dirac oscillator in the cosmic string
  spacetime}},  {\em Gen. Rel. Grav.} {\bf 45} (2013) 1847--1859,
  [\href{http://xxx.lanl.gov/abs/1307.2847}{{\tt arXiv:1307.2847}}].

\bibitem{Bakke:2011qr}
K.~Bakke and C.~Furtado, {\it {On the confinement of a quantum particle to a
  two-dimensional ring in systems described by the Dirac equation}},
  \href{http://xxx.lanl.gov/abs/1110.6458}{{\tt arXiv:1110.6458}}.

\bibitem{Hassanabadi:2015hxa}
H.~Hassanabadi, S.~Sargolzaeipor, and B.~H. Yazarloo, {\it {Thermodynamic
  Properties of the Three-Dimensional Dirac Oscillator with Aharonov?Bohm
  Field and Magnetic Monopole Potential}},  {\em Few Body Syst.} {\bf 56}
  (2015), no.~2-3 115--124.

\bibitem{VS}
A.~Vilenkin and E.~P.~S. Shellard, {\em {Cosmic strings and other topological
  defects}}.
\newblock Cambridge monographs on mathematical physics. Cambridge Univ. Press,
  Cambridge, 1994.

\bibitem{hindmarsh}
M.~Hindmarsh and T.~Kibble, {\it {Cosmic strings}},  {\em Rept.Prog.Phys.} {\bf
  58} (1995) 477--562, [\href{http://xxx.lanl.gov/abs/hep-ph/9411342}{{\tt
  hep-ph/9411342}}].

\bibitem{Katanaev:1992kh}
M.~O. Katanaev and I.~V. Volovich, {\it {Theory of defects in solids and
  three-dimensional gravity}},  {\em Annals Phys.} {\bf 216} (1992) 1--28.

\bibitem{Furtado:1994nq}
C.~Furtado and F.~Moraes, {\it {On the binding of electrons and holes to
  disclinations}},  {\em Phys. Lett.} {\bf A188} (1994) 394--396.

\bibitem{Hindmarsh:2011qj}
M.~Hindmarsh, {\it {Signals of Inflationary Models with Cosmic Strings}},  {\em
  Prog.Theor.Phys.Suppl.} {\bf 190} (2011) 197--228,
  [\href{http://xxx.lanl.gov/abs/1106.0391}{{\tt arXiv:1106.0391}}].

\bibitem{escidoc:153364}
B.~Allen and E.~P.~S. Shellard, {\it {On the evolution of cosmic strings}},  in
  {\em {The formation and evolution of cosmic strings : proceedings of a
  workshop supported by the SERC and held in Cambridge, 3-7 July, 1989}} (G.~W.
  Gibbons, S.~W. Hawking, and T.~Vachaspati, eds.), (Cambridge), pp.~421--448,
  Cambridge University Press, 1990.

\bibitem{Mota:2014uka}
H.~F. Santana~Mota and M.~Hindmarsh, {\it {Big-Bang Nucleosynthesis and
  Gamma-Ray Constraints on Cosmic Strings with a large Higgs condensate}},
  {\em Phys. Rev.} {\bf D91} (2015), no.~4 043001,
  [\href{http://xxx.lanl.gov/abs/1407.3599}{{\tt arXiv:1407.3599}}].

\bibitem{Ashour:2016cay}
A.~Ashour, M.~Faizal, A.~F. Ali, and F.~Hammad, {\it {Branes in Gravity?s
  Rainbow}},  {\em Eur. Phys. J.} {\bf C76} (2016), no.~5 264,
  [\href{http://xxx.lanl.gov/abs/1602.0492}{{\tt arXiv:1602.0492}}].

\bibitem{2016arXiv160806824F}
Z.-W. {Feng}, S.-Z. {Yang}, H.-L. {Li}, and X.-T. {Zu}, {\it {Thermodynamics
  and Phase transition of Schwarzschild black hole in Gravity's Rainbow}},
  {\em ArXiv e-prints} (Aug., 2016)
  [\href{http://xxx.lanl.gov/abs/1608.0682}{{\tt arXiv:1608.0682}}].

\bibitem{Alsaleh:2017oae}
S.~Alsaleh, {\it {Thermodynamics of rotating Kaluza-Klein black holes in
  gravity?s rainbow}},  {\em Eur. Phys. J. Plus} {\bf 132} (2017), no.~4 181,
  [\href{http://xxx.lanl.gov/abs/1704.0740}{{\tt arXiv:1704.0740}}].

\bibitem{Momeni:2017cvl}
D.~Momeni, S.~Upadhyay, Y.~Myrzakulov, and R.~Myrzakulov, {\it {Cosmic string
  in gravity?s rainbow}},  {\em Astrophys. Space Sci.} {\bf 362} (2017),
  no.~9 148, [\href{http://xxx.lanl.gov/abs/1703.0022}{{\tt arXiv:1703.0022}}].

\bibitem{birrell1984quantum}
N.~D. Birrell and P.~C.~W. Davies, {\em Quantum fields in curved space}.
\newblock Cambridge university press, 1984.

\bibitem{nakahara2003geometry}
M.~Nakahara, {\em Geometry, topology and physics}.
\newblock CRC Press, 2003.

\bibitem{greiner1990relativistic}
W.~Greiner, {\em Relativistic quantum mechanics:Wave equations}, vol.~3.
\newblock Springer, Berlin, 2000.

\bibitem{Bjorken}
J.~M. Bjorken and S.~D. Drell, {\em Relativistic Quantum Mechanics}.
\newblock McGraw-Hill Book Company, New York, 1964.

\bibitem{Abramowitz}
M.~Abramowitz and I.~Stegun, {\em Handbook of Mathematical Functions}.
\newblock Dover Publications, 1965.

\bibitem{Arfken}
G.~B. Arfken and H.~J. Weber, {\em Mathematical Methods for Physicists, sixth
  edition}.
\newblock Elsevier Academic Press, New York, 2005.

\end{thebibliography}

\end{document}